%% file: threshold_expansion.tex
\documentclass[aps,prd,onecolumn,preprintnumbers,floatfix,showpacs,
nofootinbib,superscriptaddress,notitlepage]{revtex4-1}

\usepackage{amsmath}           
\usepackage{amssymb}           
\usepackage{color}             
\usepackage{bbm}               
\usepackage{braket}            
\usepackage{xspace}            
\usepackage{mciteplus}         
\usepackage{mathtools}         
\usepackage{graphicx}          
\usepackage{slashed}           
\usepackage{enumerate}         
\usepackage[export]{adjustbox} 
\usepackage[english]{babel}    
\usepackage[utf8]{inputenc}    
\usepackage{hyperref}          
\usepackage{esint}  
\include{shortcuts}            

\hypersetup{ 
    pdfnewwindow=true,         
    colorlinks=true,           
    linkcolor=blue,            
    citecolor=blue,            
    filecolor=blue,            
    urlcolor=blue              
} 

\definecolor{jlab_red}{RGB}{192,39,45}
\definecolor{jlab_orange}{RGB}{249,102,0}
\definecolor{jlab_blue}{RGB}{47,122,121}
\definecolor{jlab_green}{RGB}{65,125,10}

\newcommand{\HSpert}[0]{Hansen:2015zta}
\newcommand{\Spert}[0]{Sharpe:2017jej}
\newcommand{\allEexps}[0]{Huang:1957im,Luscher:1986pf,Beane:2007qr,Detmold:2008gh,Hansen:2015zta,Hansen:2016fzj,Sharpe:2017jej}

\newcommand{\jlab}{Thomas Jefferson National Accelerator Facility, 
12000 Jefferson Avenue, Newport News, VA 23606, USA}
\newcommand{\odu}{Department of Physics, 
Old Dominion University, 
Norfolk, Virginia 23529, USA}
\newcommand{\cern}{Theoretical Physics Department, 
CERN, 1211 Geneva 23, Switzerland}

\begin{document}

\title{Consistency checks for two-body finite-volume matrix elements: \\
 II. Perturbative systems
}


\author{Ra\'ul A. Brice\~no}
\email[e-mail: ]{rbriceno@jlab.org}
\affiliation{\jlab}
\affiliation{\odu}

\author{Maxwell T. Hansen}
\email[e-mail: ]{maxwell.hansen@cern.ch}
\affiliation{\cern}

\author{Andrew W. Jackura}
\email[e-mail: ]{ajackura@odu.edu}
\affiliation{\jlab}
\affiliation{\odu}

\preprint{JLAB-THY-19-3113}
\preprint{CERN-TH-2020-015}

\begin{abstract}
Using the general formalism presented in Refs.~\cite{Briceno:2015tza, Baroni:2018iau}, we study the finite-volume effects for the $\2+\Jc\to\2$ matrix element of an external current coupled to a two-particle state of identical scalars with perturbative interactions. Working in a finite cubic volume with periodicity $L$, we derive a $1/L$ expansion of the matrix element through $\mathcal O(1/L^5)$ and find that it is governed by two universal current-dependent parameters, the scalar charge and the threshold two-particle form factor. We confirm the result through a numerical study of the general formalism and additionally through an independent perturbative calculation. We further demonstrate a consistency with the Feynman-Hellmann theorem, which can be used to relate the $1/L$ expansions of the ground-state energy and matrix element. The latter gives a simple insight into why the leading volume corrections to the matrix element have the same scaling as those in the energy, $1/L^3$, in contradiction to earlier work, which found a $1/L^2$ contribution to the matrix element. We show here that such a term arises at intermediate stages in the perturbative calculation, but cancels in the final result.
\end{abstract}
\date{\today}
\maketitle

\section{Introduction}\label{sec:Intro}

Understanding the emergence of hadrons from the interactions of their constituent quarks and gluons has remained a challenge, even many decades after the formulation of the fundamental theory of quantum chromodynamics (QCD). In recent years, significant progress has been made in determining the properties single-hadron ground states via numerical calculations using lattice QCD~\cite{Durr:2008zz,Fodor:2012gf,Borsanyi:2014jba}. 
Most states, however, manifest as resonances in multi-hadron scattering processes, and are rigorously defined only as poles in analytically continued scattering amplitudes.
In addition, while hadronic amplitudes allow the extraction of masses, widths and couplings, to constrain structural, information including charge radii or parton distribution functions, one must calculate and analytically continue electroweak transition amplitudes, in which an external current is coupled to the multi-hadron scattering states.

Determining scattering and transition amplitudes in lattice QCD calculations is complicated by the fact that the latter are necessarily performed in a finite Euclidean spacetime, where one cannot directly construct asymptotic states.
Presently, the most systematic method to overcome this issue is to derive and apply non-perturbative mappings between finite-volume spectra and matrix elements (which are directly calculable) and infinite-volume scattering and transition amplitudes.
This methodology was first introduced by L\"uscher~\cite{Luscher:1986pf,Luscher:1990ux}, in the context of relating the finite-volume energies of two pions, in a cubic periodic volume of length $L$, to the elastic $\2\to\2$ scattering amplitude.  

Within this framework, on-shell intermediate states yield power-law finite-volume corrections, $\Oc(1/L^{n})$, while the contribution from off-shell quantities is exponentially suppressed, scaling as $e^{-m_{\pi}L}$, where $m_\pi$ is the pion mass. For sufficiently large box sizes, the second class of corrections can be neglected, giving a systematic path towards extracting scattering observables. In the past decades, L\"uscher's formalism has been extended to include non-zero momentum in the finite-volume frame as well as coupled two-particle channels and particles with spin~\cite{Rummukainen:1995vs, Kim:2005gf, He:2005ey,  Leskovec:2012gb, Hansen:2012tf, Briceno:2012yi, Briceno:2013lba, Briceno:2014oea, Romero-Lopez:2018zyy}. 
Lattice QCD applications of the methodology have proven highly effective in the determination of two-hadron bound and resonant states \cite{Dudek:2010ew,Beane:2011sc,Pelissier:2012pi,Dudek:2012xn,Liu:2012zya,Beane:2013br,Orginos:2015aya,Berkowitz:2015eaa,Lang:2015hza,Bulava:2016mks,Hu:2016shf,Alexandrou:2017mpi,Bali:2017pdv,Bali:2017pdv,Wagman:2017tmp,Andersen:2017una,Brett:2018jqw,Werner:2019hxc,Mai:2019pqr,Wilson:2019wfr}, including those at energies where multiple channels are kinematically open~\cite{Wilson:2014cna,Dudek:2014qha,Wilson:2015dqa,Dudek:2016cru,Briceno:2016mjc,Moir:2016srx,Briceno:2017qmb,Woss:2018irj,Woss:2019hse}. 
This success in the two-hadron sector has also motivated the extension to $\2 \to \3$ and $\3\to\3$ scattering \cite{Hansen:2014eka,Hansen:2015zga,Mai:2017bge,Hammer:2017kms,Briceno:2017tce,Briceno:2018aml,Mai:2018djl,Briceno:2018mlh,Guo:2018zss,Blanton:2019igq}, with the first lattice QCD computations of the $3\pi^+$ system published last year \cite{Horz:2019rrn,Blanton:2019vdk,Mai:2019fba,Culver:2019vvu}.\footnote{For recent reviews on this topic we point the reader to Refs.~\cite{Briceno:2017max, Hansen:2019nir}.}

Extensions of these finite-volume mappings have also been derived to extract electroweak transition amplitudes from lattice QCD calculations. 
As first shown in Ref.~\cite{Lellouch:2000pv} in the context of $K\to\pi\pi$ decays, finite-volume matrix elements are related to electroweak transition amplitudes through a mapping that depends on both the box size and the scattering amplitude of the multi-particle final state. 
This has been generalized to arbitrary $\1 + \Jc \to \2$ amplitudes~\cite{Kim:2005gf,Christ:2005gi,Hansen:2012tf,Briceno:2014uqa,Briceno:2015csa,Agadjanov:2016fbd} and applied in lattice QCD studies of $K \to \pi \pi$ decay  \cite{Blum:2011ng,Boyle:2012ys,Blum:2015ywa,Bai:2015nea} as well as $\gamma^\star \to \pi \pi$ \cite{Feng:2014gba, Andersen:2018mau}
and $\pi \gamma^\star \to \pi \pi$ \cite{Briceno:2015dca, Briceno:2016kkp, Alexandrou:2018jbt} transition amplitudes. The ideas have further been generalized to $\1+ \Jc  \to \1+ \Jc$ matrix elements, with long range-contributions form multi-particle intermediate states
\cite{Christ:2015pwa,Feng:2018pdq,Briceno:2019opb}.
Most recently, the formalism for $\2+\Jc\to\2$ electroweak transition amplitudes has been developed~\cite{Briceno:2015tza, Baroni:2018iau}, generalizing previous studies based in fixed-order calculations in a specific effective field theory~\cite{Bernard:2012bi, Briceno:2012yi}. As compared to the $\1+\Jc\to\2$ methodology, the relations in Refs.~\cite{Briceno:2015tza, Baroni:2018iau} are more complicated due to the presence of additional finite-volume effects from triangle-diagram topologies. 

Because the $\2+\Jc\to\2$ finite-volume mapping is complicated, it is necessary to provide various non-trivial checks on the formalism, and calculate limiting cases in which more straightforward predictions may be extracted. 
With this in mind, in a previous study \cite{Briceno:2019nns} we provided two important checks on the general relations. First we demonstrated that the formalism, in conjunction with the Ward-Takahashi identity, protects the electromagnetic charge of finite-volume states. Though obvious from the general properties of the theory, in the context of our mapping this required exact cancellations between various combinations of finite-volume functions and thus provided a clear demonstration that all effects have been properly incorporated. 
We then explored the bound-state limit of matrix elements and recovered the expected result, that finite-volume corrections scale as $e^{- \kappa L}$ for large $L$, where $\kappa$ is the binding momentum of the two-particle bound state.  As is already well known for finite-volume bound-state energies~\cite{Beane:2003da, Davoudi:2011md, Briceno:2013bda}, in the case of a shallow bound state, $\kappa L \ll m_\pi L$, many terms in the large-volume expansion (scaling as powers of  $e^{- \kappa L}$) give significant contributions. As a result, we find it is crucial to consider the all-orders framework of Refs.~\cite{Briceno:2015tza, Baroni:2018iau} to extract reliable predictions of bound-state form factors.

In the present article, we continue our series of consistency checks by studying the $1/L$ expansion of the finite-volume matrix element, $\bra{E_0, L} \mathcal J(0) \ket{E_0,L}$, where $\ket{E_0,L}$ is the ground state of a perturbative two-scalar system and $\mathcal J(x)$ is a scalar current.  In contrast to our previous check, here we restrict attention to finite-volume scattering states meaning that the energy, $E_0(L)$, approaches the two-scalar threshold as $L \to \infty$. For finite $L$, both $E_0(L)$ and $\bra{E_0, L} \mathcal J(0) \ket{E_0,L}$ admit $1/L$ expansions with coefficients depending on the geometry of the finite volume as well as infinite-volume parameters governing the interactions. In the case of the energy, the expansion is well-known and has been studied, through various orders in $1/L$, in Refs.~\cite{\allEexps}. An analogous study for matrix elements was performed in Ref.~\cite{Detmold:2014fpa}, in which non-relativistic quantum mechanics is used to expand an $n$-particle ground-state matrix element through $1/L^4$.

In this work, we derive the $1/L$ expansion of $L^3  \bra{E_0, L} \mathcal J(0) \ket{E_0,L}$ through $\mathcal O(1/L^5)$. We compare our result with Ref.~\cite{Detmold:2014fpa} and find significant disagreement, including a difference in the behavior of the leading volume correction, with our result scaling as $1/L^3$ and that of the earlier work as $1/L^2$. To confirm our own determination, we cross-check both through a numerical study of the general formalism and through an independent perturbative calculation. In addition we use the Feynman-Hellmann theorem to relate $ \bra{E_0, L} \mathcal J(0) \ket{E_0,L}$ to a mass derivative of $E_0(L)$ and show that this enforces certain common features between the two expansions, e.g.~that both start at $\mathcal O(1/L^3)$. Finally, in our perturbative cross-check, we identify classes of terms that, if omitted, lead to the behavior reported by Ref.~\cite{Detmold:2014fpa}.

The remainder of this article is organized as follows: We first review the $1/L$ expansion of $E_0(L)$ in Sec.~\ref{sec:TE.FVS}, based in the L\"uscher scattering formalism. Then, in Sec.~\ref{sec:TE.ME}, we derive the corresponding expansion of $ \bra{E_0, L} \mathcal J(0) \ket{E_0,L}$ using the relations of Refs.~\cite{Briceno:2015tza, Baroni:2018iau}, and also describe how the results for the energy and matrix element are related via the Feynman-Hellman theorem. The main expressions are succinctly summarized in Eqs.~\eqref{eq:Eexp} and \eqref{eq:compactfinal} below. In Sec.~\ref{sec:TE.NC} we provide a numerical check of our expansion against the all-orders formalism and in \ref{sec:TE.C} we provide a detailed comparison of our result with Ref.~\cite{Detmold:2014fpa}. Section \ref{sec:PT} then describes the perturbative confirmation of our results and gives additional insight into the discrepancy with Ref.~\cite{Detmold:2014fpa}. We briefly conclude in Sec.~\ref{sec:Sum}. We also include an appendix to derive one of the technical results required for Sec.~\ref{sec:TE.ME}, concerning the imaginary part of the triangle diagram entering the infinite-volume $\2 + \Jc \to \2$ matrix element.

\section{Threshold expansion}\label{sec:TE}

In this section we review the $1/L$ expansion of the ground-state two-particle energy, $E_0(L)$, and then turn to the main result of this work, the corresponding expansion of the finite-volume matrix element. The expressions hold for a generic, relativistic quantum field theory in a periodic, cubic spatial volume with side-length $L$, provided the lowest-lying two-particle state consists of two identical scalars with mass $m$. 
We additionally require that the center-of-momentum frame (CMF) and finite-volume frame coincide, i.e.~that the particles have zero momentum, $\textbf P = \0$, in the finite volume.

For convenience we summarize the two key results here:  
\begin{enumerate}
\item In Sec.~\ref{sec:TE.FVS} we review the well-known expansion of the two-particle energy  \cite{\allEexps}
\begin{equation}
\label{eq:Eexp}
 E_0(L) =2m + \frac{4\pi a}{m L^{3}} \bigg [ 1 -  \Ic   \frac{a}{\pi L}   + \big (  \Ic^2 - \Jc \big )   \frac{a^2}{\pi^2 L^2}  + \Big ( \frac{2\pi^4  r}{a} - \frac{\pi^4 }{m^2 a^2}  - \left(  \Ic^3 - 3\Ic\Jc + \Kc  \right)   \Big)   \frac{a^3}{\pi^3 L^3}   \bigg ] + \mathcal O(1/L^7) \,,
\end{equation}
where $a$ and $r$ are the scattering length and effective range respectively, defined in Eq.~\eqref{eq:TE.ERE} below, and $m$ is the physical mass. The three geometric constants 
\begin{equation}
\mathcal I =-8.\,913\,633 \,, \qquad \mathcal J=16.\,532\,316 \,, \qquad \mathcal K =8.\,401\,924   \,,
\end{equation}
are defined and evaluated to high precision in Refs.~\cite{Luscher:1986pf,Beane:2007qr}. 
\item In Sec.~\ref{sec:TE.ME} we show that the ground-state matrix element of a scalar current at zero momentum transfer admits an analogous expansion
\begin{multline}
\label{eq:compactfinal}
L^{3} \bra{E_0,L}\Jc(0)\ket{E_0,L} = \frac{ g }{ m } \times \\ \bigg [ 1 - \frac{2\pi^4 }{m^2 a^2} \bigg (  \big (    1 -      m  a \Tc  \big )   \frac{a^3}{\pi^3 L^3}   -  \big (    1 -   2  m  a \Tc  \big )  \,  \Ic   \frac{a^4}{\pi^4 L^4} +  \big (   1 - 3 m a \Tc   \big )    \big ( \Ic^2 -  \Jc       \big )\frac{a^5}{\pi^5 L^5}  \bigg )    \bigg ] + \mathcal O(1/L^6)\,,
\end{multline}
where $\langle E_0,L \vert E_0,L \rangle =1 $,  $g$ is the scalar charge of a single particle under the scalar current, $\mathcal J(x)$, and
\begin{equation}
\label{eq:Tdef}
\mathcal T \equiv  m r +  \frac{64  \pi m^2  }{g} \Fc_0 \,.
\end{equation}
Here $\Fc_0$ is the threshold form factor, defined in Eqs.~\eqref{eq:TE.Wdf} and \eqref{eq:TE.genFF_exp} below via a straightforward relation to the infinite-volume $\2 + \Jc \to \2$ transition amplitude. 
\end{enumerate}

\subsection{Finite-volume energies}\label{sec:TE.FVS}

For a range of CMF energies from $2 m$ up to the first inelastic threshold, the finite-volume spectrum is described by the L\"uscher quantization condition~\cite{Luscher:1986pf}, which is exact up to exponentially suppressed $L$ dependence of the form $e^{- m L}$. 
The result of Ref.~\cite{Luscher:1986pf} relates the discrete energies, $E_n(L)$, to the physical scattering amplitude, by expressing the former as roots of a determinant in the space of two-particle angular momenta. 
The effects of higher angular momenta first appear in powers of $1/L$ well-beyond the orders that we control,\footnote{The leading corrections from non-trivial angular momenta enter via a finite-volume function denoted by $F_{40,00}(E,L)$ and defined, for example, in Ref.~\cite{Kim:2005gf}. This quantity encodes the mixing of the $S$-wave ($\ell=0$) and the $G$-wave ($\ell=4$) due to the reduced rotational symmetry of the cubic volume. The $F_{40,00}$-correction enters as an additive term in Eq.~\eqref{eq:TE.luscher}, scaling as $F_{40,00}(E,L)^2 = \mathcal O(1/L^8)$. The corresponding $G$-wave correction to the ground-state energy, $E_0(L)$, then scales as $1/L^{11}$ and is therefore five orders beyond the $1/L^6$ contributions that we keep.}
so that for our purposes it is sufficient to consider the truncated quantization condition
\beq\label{eq:TE.luscher}
\Mc^{-1}(E_n) = -F(E_n,L) \,,
\eeq
where $F$ is a known finite-volume function, and $\Mc$ is the $S$-wave scattering amplitude, related to the $S$-wave scattering phase shift, $\delta$, via 
\beq\label{eq:TE.M_amp}
\Mc(E) \equiv    \frac{16\pi E}{ q\cot\delta(q) - iq } \,.
\eeq
Here $q$ is the relative momentum of the two particles in the CMF, defined via $E \equiv 2\sqrt{m^2 + q^{2}}$. We recall also that $q \cot \delta(q)$ admits a convergent expansion about two-particle threshold, referred to as the effective range expansion:
\beq\label{eq:TE.ERE}
q\cot\delta = -\frac{1}{a} + \frac{1}{2} r q^{2}  +\Oc(q^{4}) \,,
\eeq
where $a$ is the scattering length and $r$ the effective range.

The finite-volume function, $F$, can be expressed in many forms, all equivalent up to exponentially suppressed corrections (see, e.g.,~Refs.~\cite{Rummukainen:1995vs,Kim:2005gf,Leskovec:2012gb,Briceno:2014oea,Briceno:2015tza}).
We begin with the following definition,
\begin{align}\label{eq:TE.fv_F}
F(E,L) & =\frac12 \lim_{\Lambda \to \infty} \bigg[     \SumInt_{\k}^{\Lambda} \bigg] \frac{1}{2\omega_{\k}} \frac{1}{E ( E - 2\omega_{\k} + i\epsilon)} \,, \\ & \equiv i\frac{  q}{16 \pi E} + F_{\pv}(E,L) \,,
\label{eq:TE.fv_F2}
\end{align}
where
\begin{equation}
\bigg[    \SumInt_{\k}^\Lambda \bigg] \equiv \frac{1}{L^{3}} \sum^{\vert \textbf k \vert < \Lambda}_{\k \in (2\pi/L) \Zbb^{3} } - \int \! \frac{ \diff^{3}\k}{(2\pi)^{3} } \, \Theta( \Lambda - \vert \textbf k \vert ) \,,
\end{equation}
and $\Theta$ is the usual Heaviside step function, included here to implement the hard cutoff.
In Eq.~\eqref{eq:TE.fv_F2} we have separated $F$ into its real and imaginary parts, denoting the former by $F_{\pv}$. The subscript ``pv'' stands for principal value, indicating that the real part of $F$ is equivalently given by taking the original definition and replacing the $i \epsilon$ pole prescription in the integral with a principal value.
Separating out the imaginary part is useful as it exactly cancels the imaginary part of the inverse scattering amplitude [see Eq.~\eqref{eq:TE.M_amp}]. It follows that Eq.~\eqref{eq:TE.luscher} is exactly equivalent to the real equation 
\begin{equation}
\label{eq:qc_real}
  q_n \cot\delta(q_n)   = -  16 \pi E_n   \, F_{\pv}(E_n,L) \,,
\end{equation}
where $q_n^2  \equiv E_n^2/4-m^2$.

From these relations, it is straightforward to determine the $1/L$ expansion of the lowest lying two-particle energy, denoted $E_0(L)$ and defined as smallest hamiltonian eigenvalue satisfying $\lim_{L \to \infty} E_0(L) = 2 m$. The infinite-volume value motivates the definitions 
\begin{align}
\label{eq:DEdef}
\Delta E_0(L) & \equiv E_0(L) - 2m \equiv 2m \bigg [ \sqrt{1 + \frac{q_0^2(L)}{m^2} } - 1 \bigg] \,, \\
& = \frac{4\pi a}{m L^{3}} \sum_{j = -2}^{\infty} \gamma_j \left( \frac{a}{\pi L} \right)^{j} \,.
\label{eq:gammadef}
\end{align}
Here the first line serves to define $\Delta E_0$ (the distance from the finite-volume state to the infinite-volume threshold) and its relation to $q_0^2$. Equation \eqref{eq:gammadef} introduces notation for a generic power series in $1/L$, and the $1/L^3$ prefactor, as well as the factors of scattering length, simplify the form of $\gamma_j$ in the final result. The aim of this subsection is to review the determination of the coefficients $\gamma_j$, defining the large-volume expansion of $E_0(L)$. 

The final non-trivial ingredient is the threshold expansion of $F_{\text{pv}}$, which can be written as 
\begin{align}\label{eq:TE.fv_F_exp}
F_{\pv}(E , L) & =  \frac{1}{4E q^{2} L^{3}}  \bigg[ 1 - \sum_{j=1}^{\infty} \left( \frac{q L }{2\pi} \right)^{2j}  \Ic_{j}  \bigg]  \,,
\end{align}
where $\Ic_j$ are numerical constants characterizing the cubic geometry,
\begin{equation}
\Ic_{j} = 
\begin{dcases} 
      \, \lim_{\Lambda \to \infty} \bigg[ \sum_{\n\ne \0}^{    n   < \Lambda} - 4 \pi \int_0^{\Lambda} \diff  n  \,  n^2 \bigg] \frac{1}{    n^2}\,, \ \  & j = 1 \,,  \\[5pt]
        \, \sum_{\n\ne \0} \frac{1  }{n^{2j}} \,,  & j \ge 2 \,,  
   \end{dcases} 
\end{equation}
with $n = \lvert \n \rvert$ and with the sums running over all non-zero integer vectors, $\n\in \Zbb^{3}/\{\textbf 0 \}$.\footnote{A convenient method to evaluate these is given in Ref.~\cite{Hansen:2016fzj}, in which an exponential damping function is used to accelerate convergence.
}

The coefficients, $\gamma_j$, can now be determined in a two step procedure: First, one substitutes the effective range expansion, Eq.~\eqref{eq:TE.ERE}, and the expansion of $F_{\text{pv}}$, Eq.~\eqref{eq:TE.fv_F_exp}, into the real version of the quantization condition, Eq.~\eqref{eq:qc_real}. In this way, both sides of the equation are expressed as polynomials in $q_0^2$ or, via the relation $q_0^2 = E_0^2/4 - m^2$, as polynomials in $E_0^2$. Second, substituting the $1/L$ expansion of $\Delta E_0(L)$ given in Eq.~\eqref{eq:DEdef}, one reaches an equality involving two series of $1/L$. The result can only be satisfied for all $L$ by tuning the values of $\gamma_j$ to enforce the equality of all coefficients. One finds $\gamma_{-2} = \gamma_{-1} = 0$, meaning that $\Delta E_0(L)$ scales as $1/L^3$. The first few non-trivial coefficients are then given by \cite{\allEexps}
\begin{gather}
\gamma_0  =  1 \,,   \qquad \qquad
\gamma_1  = -  \Ic \,,   \qquad \qquad
\gamma_2  =    \Ic^2 - \Jc  \,,    \nn \\[5pt]
\gamma_3  = - \left(  \Ic^3 - 3\Ic\Jc + \Kc  \right) +  \frac{2\pi^4  r}{a} - \frac{\pi^4 }{m^2 a^2} \,,
\end{gather}
where we have adopted the notation of Ref.~\cite{Beane:2007qr}: $\Ic_{1} = \Ic,  \  \Ic_{2} = \Jc, \  \Ic_{3} = \Kc$.\footnote{High-precision numerical determinations of these constants can also be found in that reference.}  This result is summarized in Eq.~\eqref{eq:Eexp}.

\subsection{Finite-volume matrix elements}\label{sec:TE.ME}

We now turn to the $1/L$ expansion of the finite-volume $\2+\Jc\to\2$ matrix element, where $\Jc$ is a generic scalar current density.
As above, we assume that the total momentum vanishes in the finite-volume frame, and we truncate all infinite-volume amplitudes to the $S$ wave. Then the formalism presented in Refs.~\cite{Briceno:2015tza,Baroni:2018iau}  simplifies to \beq\label{eq:TE.BH_eqn}
L^{3}\bra{E_n',L}\Jc(0) \ket{E_n,L} = \Wc_{L,\df}(E_n',E_n,L) \sqrt{\Rc(E_n',L) \Rc(E_n,L)} \,,
\eeq
where $\vert E_n, L \rangle$ is the $n$th finite-volume excited state, normalized to unity.
As with the L\"uscher quantization condition, this relation holds up to the first inelastic threshold and is exact up to exponentially suppressed corrections of the form $e^{-mL}$.

The right-hand side is composed of the Lellouch-L\"uscher factor, $\Rc$,  defined via

\begin{align}\label{eq:TE.LL_factor}
\Rc(E_n,L) & \equiv  \lim_{E \to E_n} \frac{E - E_n}{F^{-1}(E,L) + \Mc(E) } \,, \\ 
& =  - \Mc^{-2}(E_n)  \lim_{E \to E_n } \bigg [ \dfrac{\partial}{\partial E} \Big( F_{\pv}(E,L) + \dfrac{\  q}{16 \pi E} \cot\delta \Big) \bigg ]^{-1}   \,,
\end{align}
and $\Wc_{L,\df}$, a finite-volume quantity that contains the infinite-volume $\2 + \Jc \to \2$ transition amplitude \begin{align}\label{eq:TE.WLdf}
\Wc_{L,\df}(E',E,L) & = \Wc_{\df} (E',E)  + f(Q^2) \, \Mc(E') \, G(E',E,L) \, \Mc(E) \,.
\end{align}
Here $f(Q^2)$ is the single-particle form factor with momentum transfer $Q^2 \equiv -(E' - E)^2$. In the forward limit this becomes the scalar charge, denoted by $g\equiv f(0)$. In the second term in Eq.~\eqref{eq:TE.WLdf} we have also introduced $G$, a double-pole finite-volume function given explicitly by
\begin{align}\label{eq:TE.fv_G}
G(E',E,L) =  \lim_{\Lambda \to \infty} \bigg[    \SumInt^{\Lambda}_{\k} \bigg]  \frac{1}{2\omega_{\k}} \frac{1}{E'(E' - 2\omega_{\k} + i\epsilon)} \frac{1}{E (E - 2\omega_{\k} + i\epsilon)}.
\end{align}

The final ingredient in the definition of $\Wc_{L,\df}$ is the first term on the right-hand side of Eq.~\eqref{eq:TE.WLdf}, the infinite-volume divergence-free  transition amplitude, $\Wc_{\df}$.
Here `divergence free' refers to the subtraction of diagrams where the current probes one of the external legs. 
(See Refs.~\cite{Briceno:2015tza,Baroni:2018iau} for a detailed discussion of the relation between $\Wc_\df$ and infinite-volume matrix elements.)
Though the long-distance poles have been removed, $\Wc_{\df}$ does still contain two other types of kinematic singularities: (\textit{i}) threshold singularities arising from the two-particle initial and final state interactions, analogous to those in the standard $\2\to\2$ scattering amplitude, and (\textit{ii}) anomalous triangle singularities, which occur at the boundaries of the kinematic region where all intermediate states of the triangle topology can go on shell.

\bigskip

For the remainder of this article, we focus on the special case where $E' = E$, \ie~we evaluate the matrix element at zero momentum transfer. One of the many simplifying features of this limit is that the anomalous triangle singularities, type (\textit{ii}) above, then only arise at threshold and are completely given by the imaginary part of the integral defining $G(E,E,L)$. Since the sum in $G$ is pure real, this is equal (up to a minus) to 
\begin{align}\label{eq:TE.fv_G_Q0}
\text{Im} \, G(E,E,L) & = \text{Im}  \lim_{\Lambda \to \infty}  \bigg[   \SumInt_{\k}^{\Lambda} \bigg]  \frac{1}{2\omega_{\k}} \frac{1}{E^2(E - 2\omega_{\k} + i\epsilon)^2} = -   \frac{1}{32\pi E q} \,,
\end{align}
where the final equality is proven in Appendix \ref{sec:App.ImagTri}.
It will prove convenient in the following to also introduce notation for the real part of $G$. We define
\beq
G_\pv(E,L) \equiv \text{Re} \, G(E,E,L) \equiv  \lim_{\Lambda \to \infty}  \bigg[   \SumInt_{\k}^{\Lambda} \bigg]  \frac{1}{2\omega_{\k}} \frac{1}{E^2(E - 2\omega_{\k} + i\epsilon)^2} + \frac{i}{32\pi E q}  \,.
\eeq

One can remove both the usual threshold singularities ($\propto q$) and the threshold triangle singularities ($\propto 1/q$) by introducing a zero-momentum-transfer two-hadron form factor, $\Fc(E)$, related to $\Wc_{\df}(E,E)$ via
\beq\label{eq:TE.Wdf}
\Wc_{\df}(E,E) = \Mc(E) \, \left [ \Fc(E) + i   \frac{g}{32\pi E q} \right ] \, \Mc(E) \,.
\eeq
Here the $S$-wave scattering amplitude, $\Mc$, 
removes the initial- and final-state two-particle interactions so that $\Fc$ does not contain the threshold cusp appearing in $\Mc$ and $\Wc_\df$. The second term, taken directly from Eq.~\eqref{eq:TE.fv_G_Q0}, then removes the remaining singular behavior.

This completes our general discussion of the building blocks entering Eq.~\eqref{eq:TE.BH_eqn}.
Substituting the finite- and infinite-volume functions at zero momentum transfer into the general result, we deduce an all-orders expression for the finite-volume matrix element in the $S$-wave only approximation
\begin{align}
\label{eq:TE.MatrixElement}
L^{3} \bra{E_n,L}\Jc\ket{E_n,L} 
&=  \dfrac{ \vphantom{\frac{a}{b}}  \Fc(E_n) + g  \, G_{\pv}(E_n,L) }{ - \dfrac{\partial}{\partial E} \Big( 
  F_{\pv}(E,L)  
  + 
   \dfrac{1}{16\pi E} q \cot\delta 
\Big)  \Big\rvert_{E = E_n} } .
\end{align}
Given finite-volume energies and matrix elements, e.g.~computed from lattice QCD, Eq.~\eqref{eq:TE.MatrixElement} can be used to solve for the unknown $\Fc$. Together with the scattering amplitude, $\mathcal M$, and the single-particle charge, $g$, this yields a prediction for the full $\2 + \Jc \to \2$ transition amplitude in the kinematic region around the zero-momentum-transfer point.

In the present article, however, our aim is to analytically study the $L$ dependence of the threshold matrix element, $\bra{E_0,L}\Jc\ket{E_0,L} $, by expanding the right-hand side of Eq.~\eqref{eq:TE.MatrixElement} in powers of $1/L$. 
Specifically, we expand $L^3 \bra{E_0,L}\Jc\ket{E_0,L} $ through $\Oc(L^{-5})$, corresponding to four non-trivial orders in the matrix element's large volume behavior. To set up the calculation we introduce an expression analogous to Eq.~\eqref{eq:gammadef} above
\begin{equation}
\label{eq:betacoeff}
L^{3} \bra{E_0,L}\Jc\ket{E_0,L} = \frac{ g }{m } \sum_{j=0}^{\infty} \beta_j \left( \frac{a}{\pi L} \right)^{j} \,,
\end{equation}
where, as before, we have removed various factors to simplify the expressions of $\beta_j$ that arise in our final result.

We next expand all quantities entering Eq.~\eqref{eq:TE.MatrixElement} about $E^2 = 4m^2$, equivalently about $q^2 = 0$, beginning with
\beq\label{eq:TE.genFF_exp}
\Fc(E) \equiv \Fc_0 + \Oc(q^2) \,.
\eeq
We will see below that $\Fc_0$ first contributes to $L^3 \bra{E_0,L}\Jc\ket{E_0,L} $ at $\mathcal O(1/L^3)$, i.e.~to $\beta_3$, implying that the $\Oc(q^2)$ corrections first enter at $\mathcal O(1/L^6)$ ($\beta_6$) and are beyond the order  we work.
Next, the finite-volume $G$ function has a similar expansion to that given in Eq.~\eqref{eq:TE.fv_F_exp}, but with the leading $L$ scaling enhanced by the $(E - 2m)^2 \sim q^{4}$ pole in the summand. Using Eq.~\eqref{eq:double_pole} in the appendix, one can readily recover the full expansion through a derivative relation to $F_{\pv}$:
\begin{align}\label{eq:TE.fv_G_exp}
G_{\pv}(E,L) & =
-\frac{1}{E} 
\dfrac{\partial}{\partial q^2} \!
\big [
  E   F_{\pv}(E,L) \big],
 \\ 
 & =
  \frac{1}{4 E q^{4} L^{3}  } \bigg [ 1 + \sum_{j = 1}^{\infty} (j-1) \left( \frac{q L }{2\pi} \right)^{\!\!2j} \Ic_j \bigg] \,.
\label{eq:TE.fv_G_exp2}
\end{align}
Note that, when evaluated at the finite-volume ground state energy, $q^2 = \mathcal O(1/L^3)$ implying $G_{\pv} = \mathcal O(L^3)$. In Eq.~\eqref{eq:TE.MatrixElement} this leads to an $\mathcal O(L^3)$ scaling of the numerator, which is, however,  canceled by the same scaling in the denominator so that $L^3   \bra{E_0,L}\Jc\ket{E_0,L} $ is finite as $L \to \infty$.

To conclude the exercise we rewrite the denominator of Eq.~\eqref{eq:TE.MatrixElement} as a derivative with respect to $q^2$ and expand the remaining functions to reach
\begin{align}
\label{eq:TE.Matrix_exp}
L^{3} \bra{E_0,L}\Jc\ket{E_0,L} 
&=  \dfrac{ \vphantom{\frac{a}{b}}  \Fc_{0} +  
  \dfrac{g}{4 E q^{4} L^{3}  } \bigg[ 1 + \displaystyle\sum_{j = 1}^{\infty} (j-1) \left( \dfrac{q L }{2\pi} \right)^{2j} \Ic_j \bigg]
 }{ 
-    \dfrac{1}{32\pi } \dfrac{\partial}{\partial q^2} \bigg( 
\dfrac{4 {  \pi }}{q^{2} L^{3}}  \bigg[ 1 - \displaystyle\sum_{j=1}^{\infty} \left( \dfrac{q L }{2\pi} \right)^{2j}  \Ic_{j}  \bigg] 
  + 
\dfrac{1}{2} r q^2 
\bigg)  }  + \mathcal O(1/L^6) \,,\end{align}
where it is understood that $q^2$ is set to $q^2_0 \equiv E_0(L)^2/4 - m^2$ everywhere on the right-hand side. In the denominator we have also substituted the threshold expansion of $q \cot \delta$, through the order we require, and used the fact that the $q^2$ derivative annihilates the constant term.
Expanding this expression and matching to Eq.~\eqref{eq:betacoeff} yields the main result of this work: $\beta_0=1$, $\beta_1 = \beta_2 = 0$, 
\begin{align}\label{eq:TE.Matrix_exp_coeff}
\beta_3 & =  -  \frac{ 2\pi^4 }{m^2 a^2}   
+ \frac{2\pi^4  }{m a} \mathcal T \! \,, \\
\beta_4 & =   \left [  \frac{2\pi^4 }{m^2 a^2}     -  \frac{4 \pi^4  }{m a}  \mathcal T \right ] \Ic  \,, \\
\beta_5 & = \left [  -  \frac{2 \pi^4}{m^2 a^2 } +  \frac{6 \pi^4}{  m a }   \mathcal T \right ]  (\mathcal I^2 - \mathcal J) \,, \end{align}
where $\mathcal T$ is a combination of $\mathcal F_0$ and the effective range, $r$, defined in Eq.~\eqref{eq:Tdef} above. 
The leading order term, $\beta_0$, represents a pure single-hadron contribution which arises from the $G$ function, while the two-hadron form-factor $\Fc_{0}$ is sub-leading, along with relativistic corrections from the single-hadron term. 

\bigskip

We close the subsection with a simple argument that explains the absence of $1/L$ and $1/L^2$ terms, and also gives insight into the pattern of geometric constants entering $\beta_3$, $\beta_4$ and $\beta_5$. If we work in a generic scalar field theory with the field $\varphi(x)$ creating a single particle state, one possibility is to choose $\mathcal J(x) \propto \varphi(x)^2$ for the scalar current. Then, by the Feynman-Hellman theorem, the finite-volume matrix element is proportional to a mass derivative of the ground-state energy. Given the result $E_0(L) = 2m + \mathcal O(1/L^3)$, this immediately implies that $L^3 \langle E_0, L \vert \mathcal J(0) \vert E_0, L \rangle = g/m + \mathcal O(1/L^3)$, i.e.~the absence of $1/L$ and $1/L^2$ terms in the energy implies the same must hold for the matrix element. Here the factor of $L^3$, multiplying the matrix element, is required because the contribution appearing in the Hamiltonian is not directly $\mathcal J(x)$ but rather $\int_{L^{3}} \text{d}^3 \textbf x \, \mathcal J(x)$. 

Indeed, the full result can be derived from the Feynman-Hellman theorem via the relation
\begin{equation}
L^3 \langle E_0, L \vert \mathcal J(0) \vert E_0, L \rangle = g \frac{\diff E_0(L)}{\diff m^2} \,.
\end{equation}
The derivative corresponds to varying the physical mass by varying the bare mass in the Lagrangian, while keeping all other bare parameters fixed. As a result, all other physical quantities predicted by the Lagrangian inherit an $m$ dependence, while $L$ remains constant. Through the order we work one only requires an expression for the $m^2$ derivative of the scattering length. Deriving this explicitly goes beyond the scope of this article. We only note that the result \begin{equation}
  \frac{\diff a}{\diff m^2} =   \frac12 a^2 r +    \frac{32  \pi m a^2   }{g} \Fc_0   =   \frac{a^2}{2 m}   \mathcal T \,,
\end{equation}
leads to a perfect correspondence between the $1/L$ expansions of $E_0(L)$ and $L^3 \langle E_0, L \vert \mathcal J(0) \vert E_0, L \rangle $, as can be readily seen from Eqs.~\eqref{eq:Eexp} and \eqref{eq:compactfinal}.

\begin{figure*}[t!]
    \centering
\hspace*{0.9cm}
    \includegraphics[ width=0.31\textwidth]{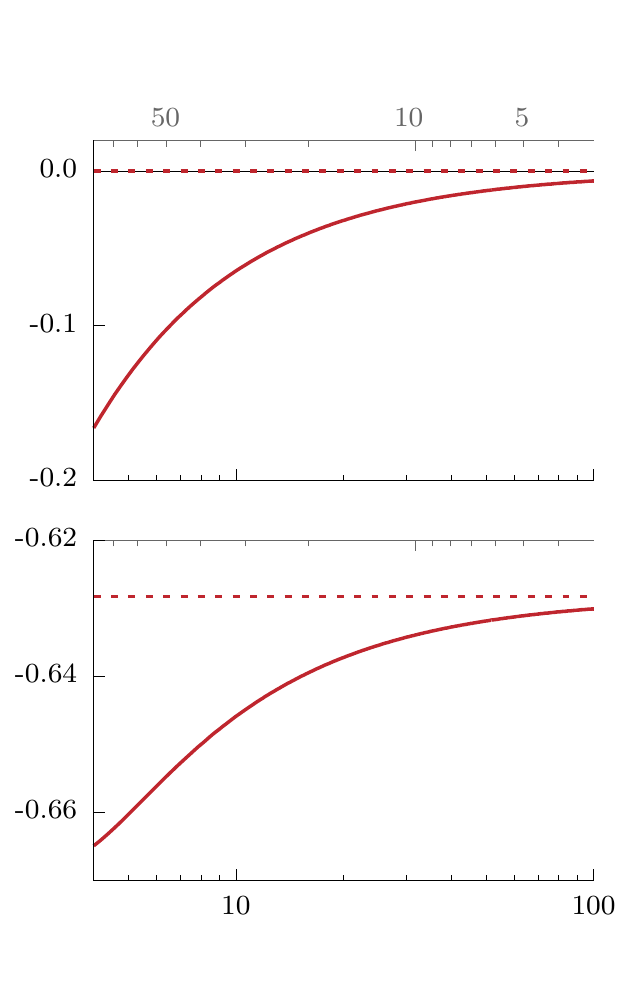}\label{fig:fig-multi_matrix_C1}
 \includegraphics[ width=0.31\textwidth]{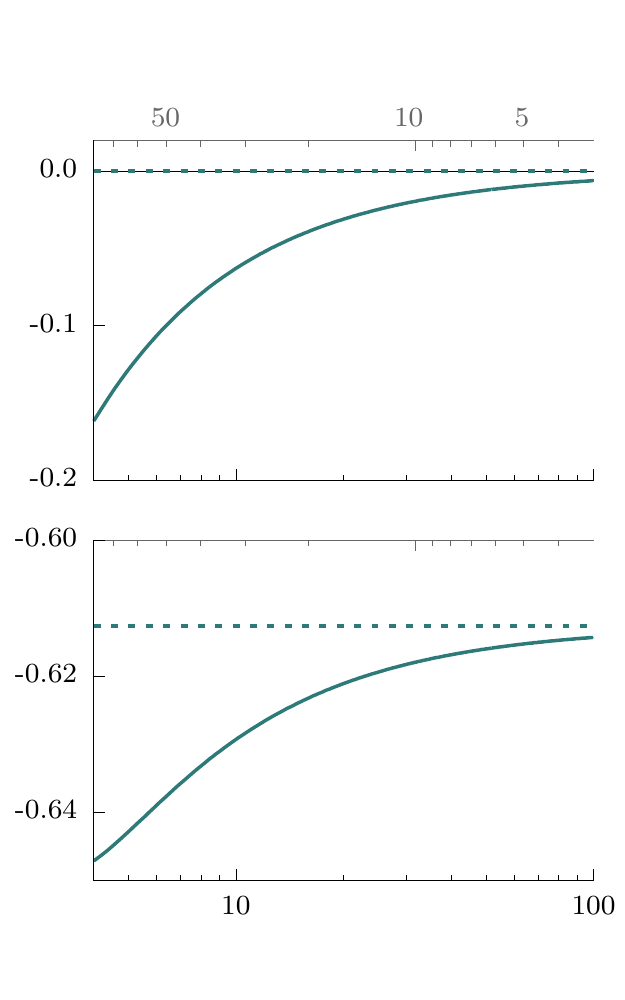}\label{fig:fig-multi_matrix_C2}
 \includegraphics[ width=0.31\textwidth]{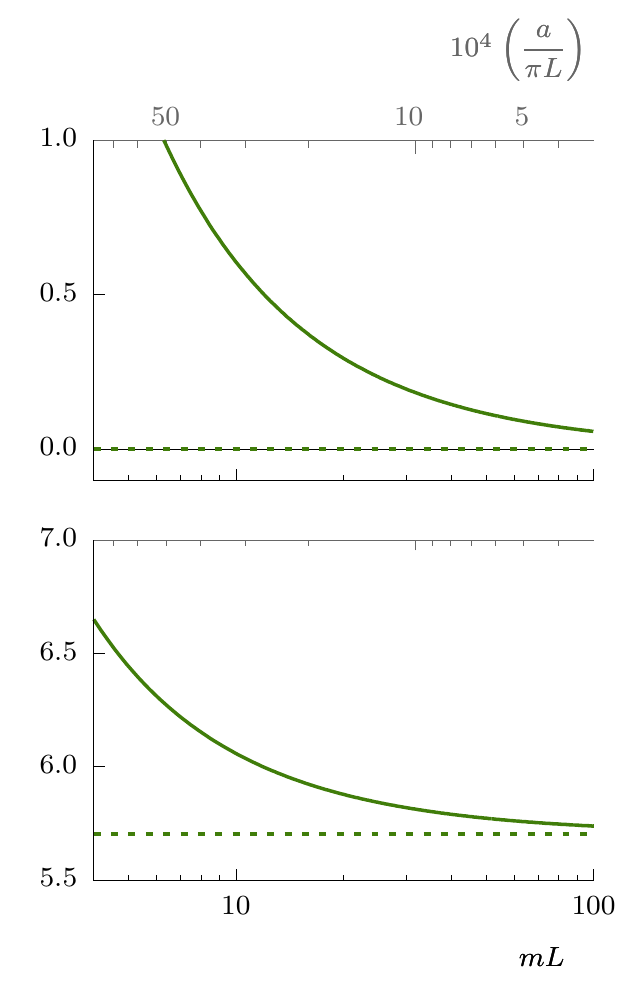}\label{fig:fig-multi_matrix_C3}
    \put(-510,150){\rotatebox{90}{\colorbox{white}{$(mL)^2 \times M_\Jc $}}}
    \put(-510,50){\rotatebox{90}{\colorbox{white}{$(mL)^3 \times M_\Jc $}}}
    \put(-440,-15){\colorbox{white}{(a) $mr = 0$,  $m\Fc_0 = 0$}}
    \put(-285,-15){\colorbox{white}{(b) $mr = 0.25$,  $m\Fc_0 = 0$}}
    \put(-125,-15){\colorbox{white}{(c) $mr = 0.25$,  $m\Fc_0 = 0.5$}}
    \caption{
Plots of $(mL)^2 \times M_{\Jc}(L)$ (top row) and $(mL)^3 \times M_{\Jc}(L)$ (bottom row) vs $mL$, with $M_{\Jc}(L)$ as defined in Eqs.~\eqref{eq:TE.MatrixElement} and \eqref{eq:TE.MJ_norm}. In each panel the solid line shows $(mL)^n \times M_{\Jc}(L)$ and the horizontal dashed line shows the expected asymptote, predicted by the analytic $1/L$ expansion. All plots are evaluated at fixed $g/m = 1.0$ and $ma = 0.1$, with $m r$ and $m \Fc_0$ varied, as indicated in the labels and in the main text.
}    \label{fig:fig-multi_matrix}
\end{figure*}

\subsection{Numerical confirmation}\label{sec:TE.NC}

To verify our strategy for expanding the general formalism in powers of $1/L$, here we numerically study the difference 
\beq\label{eq:TE.MJ_norm}
M_{\Jc}(L) \equiv  \frac{m    L^3 }{g} \bra{E_0,L} \Jc \ket{E_0,L} - \beta_0 \,,
\eeq
as a function of $mL$, using Eq.~\eqref{eq:TE.MatrixElement} to evaluate the finite-volume matrix element. By choosing various values of $ma$, $mr$, $g/m$ and $m \Fc_0$, we are able to confirm numerically that our analytic $1/L$ expansion is consistent with the general formalism. 

In Fig.~\ref{fig:fig-multi_matrix} we show the behavior of $(mL)^2 \times M_{\Jc}(L)$ (top row) and $(mL)^3 \times M_{\Jc}(L)$ (bottom row) vs $mL$, with $ma = 0.1$, $g / m = 1$, and various choices of $mr$ and $m\Fc_0$. 
In the first column we take $mr = 0$ and $m\Fc_0 = 0$, in the second $mr = 0.25$ and $m\Fc_0 = 0$, and in the third $mr = 0.25$ and $m\Fc_0 = 0.5$. 
The plots of the top row indicate that, as $mL \to \infty$, $(mL)^2 \times M_{\Jc}(L)$ asymptotes to zero, confirming the result $\beta_2  = 0$. This behavior is unchanged by varying the values of $m r$ and $m \Fc_0$, as shown. 
The plots of the bottom row show that $(mL)^3 \times M_{\Jc}(L)$ asymptotes to a non-zero value 
corresponding to $\beta_3 (ma/\pi)^3$ in the expansion. For the numerical values considered, $\beta_3 (ma/\pi)^3 = -0.63, -0.62, 5.7$, for the first, second, and third columns, respectively. 
The numerical results again confirm that there is no contribution at $\Oc(L^{-2})$ and that the first non-trivial correction, at $\Oc(L^{-3})$, is in agreement with the analytic expression for the threshold expansion. 
We have also checked that the large $L$ numerical result for the $\Oc(L^{-4})$ coincides with our expansion.

\subsection{Comparison with Ref.~\cite{Detmold:2014fpa}}\label{sec:TE.C}

In this section we compare our result, summarized in Eq.~\eqref{eq:compactfinal}, to that of Ref.~\cite{Detmold:2014fpa}, and find clear discrepancies. The earlier work uses a non-relativistic effective field theory to calculate the $1/L$ expansion for $n + \mathcal J \to n$ ground state finite-volume matrix elements, where $n$ is any number of identical scalar particles. For $n=2$ the result of Ref.~\cite{Detmold:2014fpa} becomes
\begin{equation}
\label{eq:expDetmold}
L^{3} \bra{E_0,L}\Jc\ket{E_0,L} (\textrm{Ref. \cite{Detmold:2014fpa}}) 
=
 2 \alpha_1
+
\frac{2 \alpha_1 a^2  }{\pi ^2 L^2} \mathcal J
+
\frac{\alpha_2}{L^3}
+
\frac{4 \alpha_1  a^3 }{\pi ^3 L^3} (\mathcal K-\mathcal I \mathcal J)
-
\frac{2 \alpha_2  a  }{\pi  L^4} \mathcal I
+
\frac{2  \alpha_1 a^4 }{\pi ^4 L^4}  \mathcal C + \mathcal O(1/L^5)\,,
\end{equation}
where
$\alpha_1$ and $\alpha_2$ are couplings relating the scalar current to creation and annihilation operators and $\mathcal C$ is another geometric constant, related to those specifically defined in Ref.~\cite{Detmold:2014fpa} via $\mathcal C = 3 \mathcal I^2 \mathcal J-6 \mathcal I \mathcal K-\mathcal J^2+3 \mathcal L$. The discrepancy of this result with Eq.~\eqref{eq:compactfinal} is immediately clear, in particular due to the $1/L^2$ term. As already described at the end of Sec.~\ref{sec:TE.ME}, the Feynman-Hellmann theorem implies that a $1/L^2$ correction to the matrix element requires the same for the finite-volume ground state energy. Since the latter is well-known to be absent, we are confident that this term cannot arise.

To give a more detailed comparison, we must next relate the effective-field-theory-independent parameters of our calculation, $g$ and $\mathcal T$, to the couplings that enter the earlier work. First, note that the relation between $g$ and $\alpha_1$ is given unambiguously by matching the $L \to \infty$ results of the two calculations:
  \begin{equation}\label{eq:TE.identify}
2\alpha_1(\textrm{Ref. \cite{Detmold:2014fpa}})   = \frac{g}{ m }     \,.
\end{equation}
By contrast, the expression for $\alpha_2$ is less clear. We can derive a partial relation by matching the $1/L^3$ coefficients, but it is unclear whether we should only match the $\mathcal F_0$ term within $\mathcal T$ or if we should also absorb other infinite-volume terms, e.g.~those depending on the scalar charge $g$ and scattering parameters. We take the relation
\begin{equation}
\alpha_2(\textrm{Ref. \cite{Detmold:2014fpa}})     =  g \frac{2 \pi a }{ m^3  } (m a  \mathcal T - 1) - \zeta \,,
\end{equation}
where $\zeta$ parametrizes our ignorance of the full relation and can be used to remove the $-1$ in parenthesis as well as the $r$-dependent term within $\mathcal T$. Here we do not allow the geometric constants $\mathcal I, \mathcal J$ and $\mathcal K$ to enter the relation, as these are only defined via the cubic geometry of the finite-volume, and it must be possible to relate the scattering parameters and the couplings with no reference to this.
We deduce
\begin{equation}
L^{3}\bra{E_0,L}\Jc\ket{E_0,L} (\text{this\ work})
 =
2 \alpha_1
+
\frac{\alpha_2 + \zeta}{L^3}  
-
\frac{2 ( \alpha_2 + \zeta) a  }{\pi  L^4} \mathcal I
-
\frac{4 \alpha_1 a^2  }{ m^2 L^4 } \mathcal I + \mathcal O(1/L^5)\,,
\end{equation}
Comparing to Eq.~\eqref{eq:expDetmold} we first note that the $2 \alpha_1$ and $\alpha_2/L^3$ terms now agree by construction. Thus, the only non-trivial agreement is in the $\alpha_2/L^4$ term, which exactly corresponds between the two expressions. Otherwise the results are inconsistent due to (\emph{i}) the $1/L^2$ term of Ref.~\cite{Detmold:2014fpa} and (\emph{ii}) geometric-constant-dependent discrepancies at both $\mathcal O(1/L^3)$ and $\mathcal O(1/L^4)$.

In the next section we provide a final cross-check of our result by performing an explicit perturbative calculation of the finite-volume matrix element, similar in spirit to that of Ref.~\cite{Detmold:2014fpa} but based here in a relativistic effective field theory. The results of this excercise verify our general expression and also shed light on the source of the incorrect $1/L$ scaling found in Ref.~\cite{Detmold:2014fpa}.

\section{Perturbative expansion of matrix element}\label{sec:PT}

In this section, we provide an alternative derivation of the matrix element near threshold using  perturbation theory. 
This requires deriving expansions of the finite-volume two- and three-point correlation functions, using the time-dependence to isolate the ground state, and then forming a ratio to identify $L^3 \langle E_0, L \vert \mathcal J(0) \vert E_0, L \rangle $. We work with a generalized effective field theory of a scalar field with mass $m$.

As we are interested in the two-particle threshold state, it is convenient to use an interpolator defined as the product of two scalar fields, each projected to zero spatial momentum. The two-point function is thus defined as  \beq\label{eq:2pt_corr}
C_{2\pt}(t) = \frac{(2m)^2  }{2 L^6 } e^{2imt} \braket{\tilde{\varphi}_{\0}^{2}(t) \tilde{ \varphi}_{\0}^{\dag\,2}(0) } \,,
\eeq
where $\tilde{\varphi}_{\p}(t)$ defines our notation for a single scalar field of momentum $\p$ at time $t$. This is related to the position-space and momentum-space field operators by Fourier transforms,
\begin{align}
\tilde{\varphi}_{\p}(t) & = \int_{L}\diff^3 \x \, e^{-i\p\cdot \x} \, \varphi(t,\x)  = \int \frac{\diff p_0}{2\pi} e^{-ip_0 t} \, \tilde{\varphi}(p) \,.
\end{align}
We restrict attention to $t>0$ so that we do not have to worry about time ordering, and the resultant dependence on $\vert t \vert$, for the correlation function.
The normalization of Eq.~\eqref{eq:2pt_corr} is chosen such that the correlator is unity for non-interacting limit, coinciding with the conventions chosen in Ref.~\cite{Hansen:2015zta}, with the difference that we use Minkowski time here.

The two-point correlator can be written using the usual spectral representation,
\beq
C_{2\pt}(t) = \sum_{n} Z_n e^{-i \Delta  E_n t},
\eeq
where $\Delta E_n = E_n - 2m$. Since we are only interested in the threshold state, we will explicitly isolate the $n=0$ term within $C_{2\pt}(t)$, defining
\beq\label{eq:th_spectral}
C_{2\pt,\th}(t)  = Z_0 e^{-i\Delta E_0 t} \,.
\eeq
As discussed in Ref.~\cite{Hansen:2015zta}, one can do this systematically by using the fact that excited state corrections always lead to a time dependence of the form:$\exp[-2  i ( \sqrt{m^2 + (2 \pi/L)^2 n} - m) t]$, with $n>0$. In this work we are only interested in the overlap factor $Z_0$. Following Ref.~\cite{Hansen:2015zta}, this can be determined via
\beq\label{eq:Z_corr}
Z_{0} = C_{2\pt,\th}(0) =  \frac{(2m)^2  }{2 L^6}   \big \vert  \langle 0 \vert  \tilde{\varphi}_{\0}^{2}(0)  \vert E_0, L \rangle \big \vert^2  \,.
\eeq

In a similar manner, we define the 3-point correlation function \beq\label{eq:3pt_corr}
C_{3\pt}(t',t) = \frac{(2m)^2  }{2 L^6} e^{2im(t'-t)} \braket{\tilde{\varphi}_{\0}^{2}(t') \Jc(0)  \tilde{ \varphi}_{\0}^{\dag\,2}(t)},
\eeq
where $\Jc$ is a scalar, two-field current
\beq
\label{eq:Jdef}
\Jc(x) = g \, \varphi(x)\varphi^{\dag}(x),
\eeq
and $g$ is the scalar charge.
In defining $C_{3\pt}(t',t)$ we have required $t' > 0 > t$ and have set the prefactor to match that used in the 2-point correlator. The current is renormalized in the same way as the mass-term within the Lagrangian, equivalently by requiring that $g$ is the single-hadron scalar charge to all orders.

We note here that our general result holds for any scalar current, whereas in this section we restrict attention to the single term of Eq.~\eqref{eq:Jdef}.
This is sufficient for the cross check, since the $\Jc(x)$ induces all $g$ and $\Fc_0$ terms and therefore allows one to check all terms in the general expansion. The generality is lost in this case only in that the perturbative result obscures the fact that the $1/L$ expansion, when expressed in terms of $g$ and $\Fc_0$ is universal, i.e.~holds for all scalar currents. This universality is a direct consequence of the general formalism derived in Refs.~\cite{Briceno:2015tza,Baroni:2018iau}. 

Following the procedure for the 2-point correlator, we isolate the threshold term from the spectral decomposition, giving
\begin{align}
\label{eq:C3thresh}
C_{3\pt,\th}(t',t) & =  \frac{(2m)^2  }{2 L^6}        e^{-i \Delta E_0 (t'-t)}  \langle 0 \vert  \tilde{\varphi}_{\0}^{2}(0)  \vert E_0, L \rangle      \bra{E_0,L} \Jc(0) \ket{E_0,L}   \langle E_0, L \vert \tilde{ \varphi}_{\0}^{\dag\,2}(0)   \vert 0 \rangle \,, \\[5pt]
& = Z_0 e^{-i \Delta E_0 (t'-t)} \bra{E_0,L} \Jc(0) \ket{E_0,L} \,.
\end{align}
As with the 2-point correlator, one can unambiguously separate exponentials contributing to excited states, so that this threshold correlator is straightforward to calculate, order by order in perturbation theory. The matrix element we are after is then given by the ratio
\beq\label{eq:PT.ME}
\bra{E_0,L}\Jc(0) \ket{E_0,L} = \frac{1}{Z_0} C_{3\pt,\th}(0,0) \,.
\eeq
In the following subsections, we calculate $Z_0 = C_{2\pt,\th}(0)$ and $L^3 \, C_{3\pt,\th}(0,0)$ [and thus $L^3 \bra{E_0,L}\Jc(0) \ket{E_0,L}$] through $\mathcal O(a^3, 1/L^3)$ in a generic, effective-field-theory expansion. 

We remark that the perturbative check of this section differs from the derivation of Refs.~\cite{Briceno:2015tza, Baroni:2018iau}, even though both are based in the generic properties of relativistic field theory. The key distinction is that the ground state matrix element is identified here through terms with time dependence of the form $e^{- i 2 m t}$, corresponding in momentum space to the lowest lying non-interacting finite-volume pole. Of course, the full correlator has a time dependence dictated by the interacting spectrum. This corresponds to the interacting pole positions (and the cancellation of non-interacting poles) that was identified after the all orders summation in Refs.~\cite{Briceno:2015tza, Baroni:2018iau}. 

The distinction leads to important technical differences in the calculation. In particular, in Refs.~\cite{Briceno:2015tza, Baroni:2018iau} we found that diagrams in which the current couples to a final single-particle [see Figs.~\ref{fig:Feynman_3pt}(b1-c2)] did not contribute to the residue of interacting poles that defined the matrix element of interest. In the present calculation, by contrast, these diagrams appear at the fixed-order being considered, and turn out to be necessary in recovering Eq.~\eqref{eq:compactfinal}.

\subsection{Two-point correlator}\label{sec:PT.2pt}

The order-by-order calculation of $Z_0$ [through $\mathcal O(a^3, 1/L^6)$] is one of the central ingredients in perturbative determinations of the ground state two-particle energy, described in detail in Refs.~\cite{\HSpert,\Spert}. As illustrated in detail in Ref.~\cite{\Spert}, one of the central complications in the fixed-order calculation is that numerous contribution arise that either cancel in the final result or else are absorbed in the relation between the bare coupling and the scattering length. To avoid these complications, here we present a new method, in which $Z_0$ is derived through the expansion of finite-volume correlator expressed via standard identities that arise in the context of finite-volume quantization conditions.

We begin with
\begin{align}
C_{2\pt}(t) 
& =  \frac{(2m)^2 }{2 L^6} e^{2imt} \int \frac{\diff E'}{2\pi} \int \frac{\diff E}{2\pi} \int \frac{\diff k_0'}{2\pi} \int \frac{\diff k_0}{2\pi}  \, e^{-iE' t} \,  G_L(E',E,k_0',k_0) \,, 
\end{align}
where
\begin{equation}\label{eq:PT.FV_FT_mom}
G_L(E',E,k_0',k_0) \equiv \int_L \diff^4 x \, e^{  i k'  x } \,  \int_L \diff^4 y \, e^{  i (P' - k')  y }  \,  \int_L \diff^4 z \, e^{ - i k  z }  \,  \int_L \diff^4 w \, e^{-  i (P-k)  w } \ \langle \varphi(x) \varphi(y) \varphi^\dagger(z) \varphi^\dagger(w) \rangle_L  \,,
\end{equation}
with $k^\mu = (k^0, \boldsymbol 0)$, $k'^\mu = (k'^0, \boldsymbol 0)$, $P^\mu = (E, \boldsymbol 0)$, $P'^\mu = (E', \boldsymbol 0)$. We stress here that the four-point function also includes the disconnected contractions. In addition, we note that $G_L$ is proportional to an energy conserving Delta function, $\delta(E-E')$.

We next note that, following the Lehmann-Symanzik-Zimmermann reduction formula, the connected part of $G_L$ will contain a quadruple pole of the form $[(k^2 - m^2)(k'^2 - m^2)((P-k)^2 - m^2)((P'-k')^2 - m^2)]^{-1}$ and, after projecting to zero spatial momentum, this leads to poles at $k_0, k_0' = \pm m \mp i \epsilon$. Evaluating the $k_0$ and $k_0'$ integrals by enciricling these, we find
\begin{align}
C_{2\pt, \th}(t) 
& =1 - \frac{ 1}{2 L^3} e^{2imt}  \ointclockwise_{2m} \frac{\diff E}{2\pi}    \, e^{-iE t} \,   \frac{1}{E^2}   \frac{i \mathcal M_{L}(E)}{  ( E - 2 m  + i \epsilon)^2} \,, 
\end{align}
where we have also used the Dirac delta function to evaluate the $E'$ integral. We define $i \mathcal M_{L}(E)$ as the connected and amputated part of $G_L$, evaluated at $k_0' = k_0 = m$.
Two additional comments are in order here: First, the leading $1$ arises from the disconnected pieces which exactly cancel the prefactor, by construction. Second, we have neglected higher singularities in the $k_0$ and $k_0'$ dependence, as these ultimately lead to excited state exponentials. For this reason the ``$\text{th}$'' subscript has also been added at this stage, together with the labels on the integral that indicate only pole near $E=2m$ is to be included.

The next step is to substitute the relation
\begin{equation}\label{eq:PT.FV_ML}
\mathcal M_{L}(E) = \bigg [ \frac{ q \cot \delta(q)}{16 \pi E} + F_{\text{pv}}(E,L) \bigg ]^{-1} \,,
\end{equation}
together with the definitions
\begin{equation}
\mathcal K(E) \equiv  \frac{16 \pi E}{  q \cot \delta(q)} \,, \qquad \qquad  F_{\text{pv}}(E,L) \equiv \frac{1}{4 mL^3} \frac{1}{E(E - 2m + i\epsilon)} + F'(E,L)  \,,
\end{equation}
to reach
\begin{align}
C_{2\pt, \th}(t) 
& =1 - i \frac{ 1}{2 L^3} e^{2imt}  \ointclockwise_{2m} \frac{\diff E}{2\pi}    \, e^{-iE t} \,   \frac{1}{E^2}   \frac{1}{  ( E - 2 m  + i \epsilon)^2}  \bigg [  \mathcal K(E)^{-1} +\frac{1}{4 mL^3} \frac{1}{E(E - 2m + i\epsilon)} + F'(E,L)  \bigg ]^{-1} \,.
\end{align}
The introduction of $F'(E,L)$ is motivated by the fact that this object is analytic near $E = 2m$ whereas $F_{\text{pv}}(E,L)$ contains the simple pole that we are displaying explicitly. 

Expanding the square-bracketed function in powers of $1/(E - 2m + i \epsilon)$ and evaluating the contour integral leads to the elegant result
\begin{align} 
C_{2\pt, \th}(t) 
& =  1- \frac{ 1}{2 L^3} e^{2imt}  \sum_{n=0}^\infty  \bigg (  \frac{- 1}{4mL^3} \bigg )^n     \frac{1}{(n+1)!} \frac{\partial^{n+1}}{\partial E^{n+1}} \bigg [      \frac{e^{-iE t}}{E^{2+n}} \frac{1}{ \mathcal K(E)^{-1}  + F'(E,L)}  \bigg ]_{E=2m} \,.
\end{align}
This provides a powerful tool for identifying terms in the expansion of $Z_0$ as well as the corresponding energy. Since we are only interested in the latter here, we set $t=0$ to reach the general result
\begin{equation} \label{eq:generalZ0final}
Z_0 = 1-  \frac{ 1}{2 L^3}   \sum_{n=0}^\infty  \bigg (  \frac{- 1}{4mL^3} \bigg )^n     \frac{1}{(n+1)!} \frac{\partial^{n+1}}{\partial E^{n+1}} \bigg [      \frac{1}{E^{2+n}} \frac{1}{ \mathcal K(E)^{-1}  + F'(E,L)}  \bigg ]_{E=2m} \,.
\end{equation}

\begin{figure*}[t!]
    \centering
    \includegraphics[ width=0.9\textwidth]{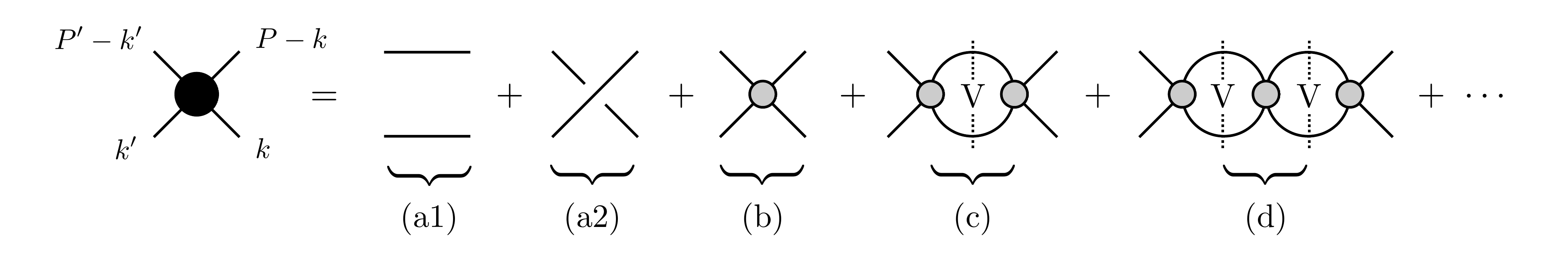}
    \caption{ Expansion of the momentum-space finite-volume two-point correlator, Eq.~\eqref{eq:PT.FV_FT_mom}, in terms of the infinite-volume scattering amplitude $\Mc$ (gray circle) and finite-volume cuts $F$ (dotted line).
The geometric series in $\Mc$ and $F$ yields $\Mc_{L}$, given in Eq.~\eqref{eq:PT.FV_ML}.
}
    \label{fig:Feynman_2pt}
\end{figure*}

The final step is to substitute the effective range expansion as well as the large $L$ expansions of $F'$:
\begin{align}\label{eq:PT.FV_thresh}
F'(2m,L)   & = - \frac{1}{32 \pi^2 mL}\Ic  + \mathcal O(1/L^3) \,,   \\
\frac{\partial}{\partial E} F'(E,L) \Big\rvert_{E = 2m}  & = -\frac{  L}{128 \pi^4} \Jc   + \mathcal O(1/L) \,,     \\
\frac{\partial^2}{\partial E^2} F'(E,L) \Big\rvert_{E = 2m}  & =- \frac{  m L^3}{256 \pi^6} \Kc +  \mathcal O(L) \,.
\end{align}
Substituting in Eq.~\eqref{eq:generalZ0final} and then re-expanding in $1/L$, we conclude
\beq\label{eq:PT.Z_2pt}
Z_0 = 1 - \left( \frac{a}{\pi L} \right)^2 \Jc  -  \frac{2 \pi^4}{m^2 a^2}  \big (1 -   m^2 a r    \big ) \left( \frac{a}{\pi L} \right)^3   - 2 (\Kc - \Ic\Jc )  \left( \frac{a}{\pi L}\right)^3    + \Oc(L^{-4}) \,.
\eeq
The first sub-leading correction at $\Oc(L^{-3})$ is due to the leading-order amplitude Fig.~\ref{fig:Feynman_2pt} (b), whereas the correction at $\Oc(L^{-2})$ arises from volume enhancements in the one-loop diagram, Fig.~\ref{fig:Feynman_2pt} (c).

\subsection{Three-point correlator}\label{sec:PT.3pt}

We now turn to the three-point correlator, which can be written as
\begin{align}\label{eq:2Jto2_corr}
C_{3\pt}(t',t) = \frac{(2m)^2  }{2 L^6} e^{2im(t'-t)} 
\int\!\frac{\diff E'}{2\pi} \int\!\frac{\diff E}{2\pi}\int\!\frac{\diff k_0'}{2\pi}\int\!\frac{\diff k_0}{2\pi} e^{-iE' t' + i E t} \braket{\tilde{\varphi}(P'-k')\tilde{\varphi}(k') \Jc(0) \tilde{\varphi}(P-k)\tilde{\varphi}(k)}  \,, \end{align}
where all four-momenta have a vanishing spatial component. As in the preceding section, we express the momentum-space correlator on the right-hand side of Eq.~\eqref{eq:2Jto2_corr} order by order in Feynman diagrams, and then calculate the contribution of each to $C_{3\pt,\th}(0,0)$, defined in Eq.~\eqref{eq:C3thresh}.

\begin{figure*}[t!]
    \centering
    \includegraphics[ width=0.9\textwidth]{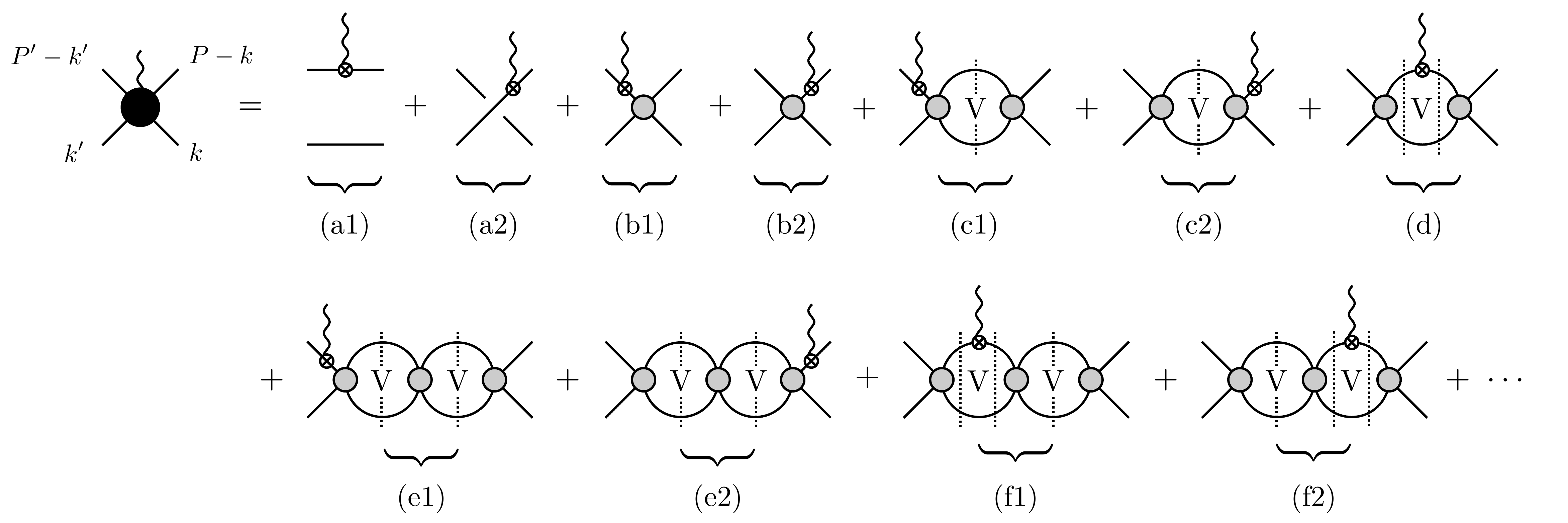}
    \caption{Expansion of the momentum-space finite-volume three-point correlator in terms of the infinite-volume scattering amplitude $\Mc$, finite-volume cuts $F$, and the $G$ function (double dotted line). 
Since the particles are identical, diagrams (a-c), and (e) contain a multiplicity $\times 2$ to account for the current probing both the upper and lower legs. 
Triangle diagrams in (d) and (f) do not include a multiplicity as there is only a single contribution for identical particles. }
    \label{fig:Feynman_3pt}
\end{figure*}

The LO contribution arises from diagrams (a1) and (a2) of Fig.~\ref{fig:Feynman_3pt}, with each diagram having a multiplicity $\times 2$ to account for the current coupling to either particle. This leads to \begin{align}
C_{3\pt,\th}^{(0)}(0,0) & =\frac{(2m)^2 }{2 L^6}   \left( 4\times \frac{g}{2 m} L^3 \right) \ointclockwise_{2m} \frac{\diff E'}{2\pi} \ointclockwise_{2m} \frac{\diff E}{2\pi} \,  \frac{i}{E'(E' - 2m + i\epsilon)}  \frac{i}{E(E - 2m + i\epsilon)}  \,, \\
& =  \frac{g}{m L^3} \,,
\end{align}
where the upstairs factor of $L^3$ comes from the momentum-conserving delta function associated with the disconnected propagator.
Alternatively, this same result is reached using propagators in the time-momentum representation
\begin{equation}
C_{3\pt,\th}^{(0)}(t',t)   = \frac{(2m)^2 }{2 L^6} e^{2im(t'-t)}  \bigg[  4 \bigg (L^3 \frac{e^{- i m t'}}{2m}  \bigg )  \frac{g}{L^6} \bigg (L^3 \frac{e^{i m t}}{2m}  \bigg ) \bigg (L^3 \frac{e^{-  i m (t'-t)}}{2m}  \bigg ) \bigg ]= \frac{g}{m L^3}   \,,
\end{equation}
where in this case the current is rewritten as $\mathcal J(0) = [g/L^6] \sum_{\textbf k, \textbf p} \tilde \varphi(0,\textbf k)  \tilde \varphi^\dagger(0,\textbf p)$, leading to the volume factor as shown. The complete leading-order calculation also includes a term in which the current is disconnected from both propagators. However this term, like every other contribution with the current fully disconnected, is cancelled by a counterterm, chosen to enforce $g$ as the physical value of the scalar charge. For this reason we omit current-disconnected diagrams throughout.

The contribution at NLO is given by the diagrams in Figs.~\ref{fig:Feynman_3pt} (b1) and (b2), where the current couples to a single hadron in the final state. 
As mentioned above, these do not contribute to the all orders derivation of Ref.~\cite{Briceno:2015tza,Baroni:2018iau} but must be included in this fixed-order calculation. The two diagrams give the same contribution to the threshold matrix element and we find
\begin{align}
C_{3\pt,\th}^{(1)} (t',t) & =   - 4 i  g  \, \mathcal M_{2,\text{th}} \,   \frac{ (2m)^2  }{2 L^6} e^{2im(t'-t)}  \frac{1}{(2m)^2}
\ointclockwise_{2m}\frac{\diff E'}{2\pi} \ointclockwise_{2m} \frac{\diff E}{2\pi}  \frac{i e^{-iE' t' + i E t}   }{E^2 (E - 2 m + i \epsilon)^2E' (E' - 2 m + i \epsilon) }  \,, \\
& =   - 4 i  g  \, \mathcal M_{2,\text{th}} \,   \frac{ (2m)^2  }{2 L^6} e^{- 2im t}  \frac{1}{(2m)^3}
  \ointclockwise_{2m} \frac{\diff E}{2\pi}  \frac{e^{ i E t}  }{E^2 (E - 2 m + i \epsilon)^2   }  \,, \\
  & =   4   g  \, \mathcal M_{2,\text{th}} \,   \frac{ (2m)^2  }{2 L^6} e^{- 2im t}  \frac{1}{(2m)^3}
 \frac{\partial}{\partial E} \frac{e^{ i E t}  }{E^2     } \bigg \vert_{E= 2m} \,,
\label{eq:C3NLO}
\end{align}
where $ \mathcal M_{2,\text{th}} = - 32 \pi m a$ is the threshold scattering amplitude.
Setting $t=t'=0$ then yields
\begin{equation}
C_{3\pt,\th}^{(1)} (0,0)   = -   \frac{g}{m L^3}  \frac{4 \pi^4}{m^2 a^2} \left( \frac{a}{\pi L} \right)^3 \,.
\end{equation}
As with the leading-order result, this can also be reproduced using time-momentum perturbation theory.

It is instructive to already collect the results for $C_{3\pt,\th} (0,0) $ and $Z_0$, through $\mathcal O(a)$. We find
\begin{equation}
\label{eq:failedcancel}
\bra{E_0, L} \mathcal J(0) \ket{E_0, L} = \frac{C_{3\pt,\th} (0,0) }{Z_0} = \frac{g}{m L^3}  \dfrac{1 - 2 \times  \dfrac{2 \pi^4}{m^2 a^2} \left( \dfrac{a}{\pi L} \right)^3 }{1 -   \dfrac{2 \pi^4}{m^2 a^2} \left( \dfrac{a}{\pi L} \right)^3 }  + \mathcal O(a^2) \,.
\end{equation}
Note that the factor of $2$ in the numerator spoils the cancellation, so that an $\mathcal O(a/L^3)$ term does contribute to the final result. This may seem surprising since all contributions considered so far involve the current coming to one of the external legs. Thus for each term in $C_{3\pt,\th} (0,0)$ one expects a closely related contribution to $Z_0$. The key point is that the relative combinatoric factors differ between the LO and NLO terms and this leads to an NLO term surviving in the matrix element. 

We now turn to contributions scaling as $a^2/L^2$. As with the NLO contributions, here the contribution to $C_{3\pt,\th} (0,0) $ with the current on the external leg does not cancel the analogous contribution to $Z_0$. However, an additional term with the current on an internal leg, does cancel against the remainder so that no $1/L^2$ behavior enters the final matrix element.
The relevant expressions arise from evaluating the next-to-next-to-leading-order (N2LO) diagrams of Figs.~\ref{fig:Feynman_3pt}(c1), (c2) and (d). The first two of these give
\begin{equation}
 C_{3\pt,\th}^{(2,c)}(0,0) =  - 2 \frac{g}{m L^3} \left( \frac{a}{\pi L} \right)^2 \Jc + \mathcal O(1/L^3) \,,
\end{equation}
and thus follow the pattern of the NLO diagram, including the factor of $2$ that spoils the complete cancellation.

The diagram of Fig.~\ref{fig:Feynman_3pt}(d) contributes a similar term as can be seen from rewriting the expression as 
 \begin{align}
 \label{eq:C3ptdStep1}
C_{3\pt,\th}^{(2,{\rm d})} (0,0)  & = - \frac{(2m)^2  }{2 L^6} \frac{g}{(2m)^2} \int\!\frac{\diff E'}{2\pi} \int\!\frac{\diff E}{2\pi} \frac{          \Mc(E')  G (E',E,L) \Mc(E)      }{E'(E' - 2m + i\epsilon) E(E - 2m + i\epsilon)}   +\mathcal O(1/L^3)  \,.
\end{align}
 Here we have introduced $G(E',E,L)$ in the perturbative expansion by rewriting the summed loop as an integral plus a sum-integral difference. The latter contributes at $1/L^3$ and is thus dropped to illustrate the leading behavior first. We can further simplify this by dividing $G(E',E,L)$ into real and imaginary parts and splitting the real part into the double pole at threshold ($\sim 1/[(E - 2m)(E'-2m)]$), together with the sum over $\textbf k \neq 0$, denoted by $G'(E',E,L)$. As with $F'(E,L)$, this term is regular near $E=E'=2m$ and the resulting contribution comes from encircling the poles shown explicitly in Eq.~\eqref{eq:C3ptdStep1}. Substituting  
\begin{equation}
G'(2m,2m,L) = \frac{L}{128 \pi^4 m } \Jc + \mathcal O(1/L) \,,
\end{equation}
together with the threshold scattering amplitude then gives
\begin{align}
\label{eq:C3ptd}
C_{3\pt,\th}^{(2,d)} (0,0) &  =  \frac{g}{m L^3} \left( \frac{a}{\pi L} \right)^2 \Jc + \mathcal O(1/L^3)   \,.
\end{align}
Alternatively, this same result can be derived in time-momentum perturbation theory. In this case one finds that the relevant time-dependence arises only for $\textbf k \neq \textbf 0$. This gives
\begin{align}
C_{3\pt,\th}^{(2,d)} (0,0) & =  g \, \mathcal M_{2,\text{th}}^2 \, \frac{(2m)^2 }{2 L^6}  \frac{1}{L^3}  \sum_{\textbf k \neq \textbf{0}}     \frac{1}{(2m)^4 (2 \omega_{\textbf k})^3   (2 \omega_{\textbf k} - 2 m)^2}  \,,
\end{align}
which is equivalent to Eq.~\eqref{eq:C3ptd} above. Combining $C_{3\pt,\th}^{(2,c)} (0,0)$ and $C_{3\pt,\th}^{(2,d)} (0,0)$ gives the same $1/L^2$ dependence as in $Z_0$, such that these terms perfectly cancel in the ratio. This is our final confirmation that $1/L^2$ scaling is absent from the matrix element: $L^3 \bra{E_0,L} \mathcal J(0) \ket{E_0,L}$.

\begin{figure*}[t!]
    \centering
    \includegraphics[ width=0.6\textwidth]{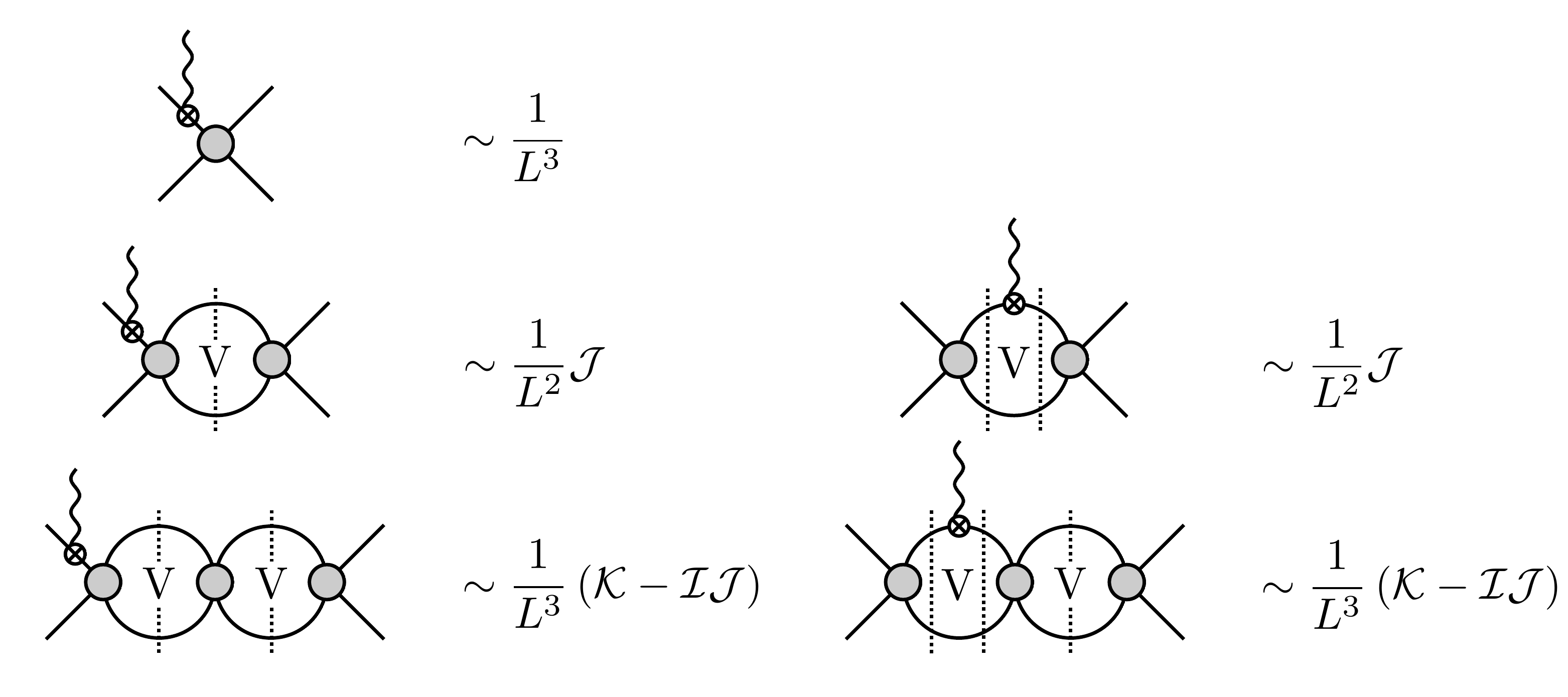}
    \caption{Leading finite-volume scaling for diagrams contributing to the matrix element from the three-point correlator. 
Diagrams on the left have identical scaling to those of the corresponding two-point correlator of the same topology, i.e.~with the external current removed.}
    \label{fig:Feynman_3pt_L_dep}
\end{figure*}

To conclude the perturbative check, we have evaluated $C_{3\pt,\th}(0,0)$ to one higher order in both $a$ and $1/L$. 
This follows the same pattern of the calculation so far, but induces $\Fc_0$ and $r$ dependent contributions as well as various geometric constants. 
One finds  
\beq \label{eq:PT.3pt_L}
\frac{m  L^3}{g} C_{3\pt,\th}(0,0) = 1 - \left( \frac{a}{\pi L} \right)^2 \Jc - \frac{4 \pi^4}{m^2 a^2}  \big (1 -    m a  \mathcal T \big ) \left( \frac{a}{\pi L} \right)^3 - 2 (\Kc - \Ic \Jc)  \left(  \frac{a}{\pi L} \right)^3    + \Oc(L^{-4}) \,.
\eeq
Combining this with Eq.~\eqref{eq:PT.Z_2pt} gives an expansion through $1/L^3$ that is completely consistent with the result of Eq.~\eqref{eq:compactfinal}.
Figure \ref{fig:Feynman_3pt_L_dep} shows the leading $1/L$ scaling for each diagram contributing to the matrix element.

We speculate that the disagreement with Ref.~\cite{Detmold:2014fpa} arises from the earlier work omitting the $1/L$ corrections to $Z_0$ and also dropping contributions to $C_{3\pt,\th}(0,0)$ with the current attached to an external leg. Equivalently, the earlier work may have been based in the assumption that the two sets of terms cancel, as they would if it were not for the leading-order discrepancy, summarized in Eq.~\eqref{eq:failedcancel}. Indeed, we find that if we drop diagrams where the current probes the external legs [(b), (c), and (e) of Fig.~\ref{fig:Feynman_3pt}], and also drop $Z_0$, then we exactly recover the $1/L$ expansion of Ref.~\cite{Detmold:2014fpa}. We stress however that there is no theory nor limiting case where this result holds and, in particular, the absence of the $1/L^2$ is a universal result inherited from the ground-state energy.

\section{Summary}\label{sec:Sum}

Understanding the structure of strongly interacting resonances and bound states requires knowledge of two-hadron electroweak transition amplitudes. 
With this in mind, a framework was presented in Refs.~\cite{Briceno:2015tza,Baroni:2018iau} to relate finite-volume matrix elements, which can be computed using lattice QCD, to infinite-volume $\2+\Jc\to\2$ transition amplitudes. 
To gain confidence in this formalism, we have performed a series of consistency checks, presented in Ref.~\cite{Briceno:2019nns} together with the present article. 
While Ref.~\cite{Briceno:2019nns} is concerned with the volume-independence of the charge and the the finite-volume effects on bound-state matrix elements, this work is dedicated to the $1/L$ expansion of the lowest-lying two-hadron scattering state.

Specifically, in Sec.~\ref{sec:TE.ME} we have derived the $1/L$ expansion of $L^3 \langle E_0,L \vert \mathcal J(0) \vert E_0, L \rangle$ through $\mathcal O(1/L^5)$, with the main result is summarized in Eq.~\eqref{eq:compactfinal}. We have confirmed that the expression matches expectations from the Feynman-Hellmann theorem, which can be used to draw a correspondence to the $1/L$ expansion of $E_0(L)$, and also agrees with an independent perturbative check. We have also compared to Ref.~\cite{Detmold:2014fpa}, in which the authors consider the $1/L$ expansion of $n + \Jc \to n$ finite-volume matrix elements, through $\mathcal O(1/L^4)$, in the context of non-relativistic quantum mechanics. For $n=2$, the results are expected to agree, since relativistic effects first appear at $\mathcal O(1/L^6)$. 
However, we find clear disagreement with the earlier publication, both in the scaling of the leading $1/L$ correction ($1/L^2$ in Ref.~\cite{Detmold:2014fpa} and $1/L^3$ in this study) and the coefficients for the sub-leading terms. In the perturbative calculation presented in Sec.~\ref{sec:PT}, we have identified classes of corrections that, if omitted, lead to the expressions found in Ref.~\cite{Detmold:2014fpa}.

\section{Acknowledgements}
We thank Will Detmold for useful discussions.
R.A.B. is supported in part by USDOE grant No. DE-AC05-06OR23177, 
under which Jefferson Science Associates, LLC, manages and operates Jefferson Lab.
R.A.B. also acknowledges support from the USDOE Early Career award, contract de-sc0019229.

\appendix

\section{Imaginary part of triangle diagram}\label{sec:App.ImagTri}

In this appendix we demonstrate that $G(E,E,L)$ has a simple imaginary part, given by Eq.~\eqref{eq:TE.fv_G_Q0} of the main text. The imaginary part arises only from the integral part of $G(E,E,L)$, and thus, the quantity we are after is given by
\begin{align}\text{Im} \, G(E,E,L) & = - \text{Im} \int^{\Lambda} \! \! \! \! \frac{\diff^3 \textbf{k}}{(2 \pi)^3}\,  \frac{1}{2\omega_{\k}} \frac{1}{E^2(E - 2\omega_{\k} + i\epsilon)^2} \,.
\end{align}
Here we have included the hard cutoff, $\Lambda$, since the real part of the integral has an ultraviolet divergence that cancels with that of the sum in $G(E,E,L)$. As we will see, the imaginary part is ultraviolet-finite and therefore also universal.

Next it is convenient to expand the integrand about the douple pole at $q^2 = k^2$, where $q^2 = E^2/4-m^2$ and $k^2 = \textbf k^2$,
\beq
\label{eq:double_pole}
\frac{1}{ 2\omega_{\k} } \frac{1}{ E^2 (E - 2\omega_{\k} + i\epsilon)^2} = \frac{1}{2\omega_{\k}} \frac{(E + 2\omega_{\k})^2}{ (4 E)^2 (q^2 - k^2 + i\epsilon)^2}  = \frac{1}{4E} 
\frac{1}{(q^2 - k^2 + i\epsilon)^2} + \Oc\left [ \left(k^2 - q^2 \right)^{\! 0} \right ]\,.
\eeq
This is useful because the sub-leading terms only contribute to the real part of the integral. We reach
\beq\label{eq:App.Ginf_q2}
\text{Im} \, G(E,E,L)   = - \frac{1}{8 \pi^2 E}  \int_{0}^{\infty} \diff k \frac{k^2}{(q^2 - k^2 + i\epsilon)^2}  \,,
\eeq
where we have also used that the singular piece gives a convergent integral so that we can send $\Lambda \to \infty$.
To identify the imaginary part, we rewrite Eq.~\eqref{eq:App.Ginf_q2} as a contour integral,
\beq
 \int_{0}^{\infty} \diff k \frac{k^2}{(q^2 - k^2 + i\epsilon)^2} = \frac{1}{2} \int_{-\infty}^{\infty} \diff k \frac{k^2}{(q^2 - k^2 + i\epsilon)^2} = \frac{1}{2} \oint \diff k \frac{k^2}{(k - q - i\epsilon)^2(k + q + i\epsilon)^2},
\eeq
where, for concreteness, we envision closing the contour in the upper-half plane. 
Evaluating the integral, we pick up the residue at the pole, $k =   q + i\epsilon$,
\beq
 \oint \diff k \frac{k^2}{(k - q - i\epsilon)^2(k + q + i\epsilon)^2} = 2\pi i \frac{\diff}{\diff k} \frac{k^2}{(k + q)^2} \bigg\rvert_{k = q} =  i \frac{\pi}{2q} \,,
\eeq
and thereby conclude the desired result
\beq
\text{Im} \, G(E,E,L)  =  -   \frac{1}{32\pi E q} \,.
\eeq

\bibliography{bibi}     

\end{document}

%% file: shortcuts.tex
\renewcommand{\k}{\mathbf{k}}
\newcommand{\p}{\mathbf{p}}

\newcommand{\x}{\mathbf{x}}

\newcommand{\0}{\mathbf{0}}
\newcommand{\1}{\mathbf{1}}
\newcommand{\2}{\mathbf{2}}
\newcommand{\3}{\mathbf{3}}

\newcommand{\n}{\mathbf{n}}

\newcommand{\Zbb}{\mathbbm{Z}}

\newcommand{\Fc}{\mathcal{F}}

\newcommand{\Ic}{\mathcal{I}}
\newcommand{\Jc}{\mathcal{J}}
\newcommand{\Kc}{\mathcal{K}}

\newcommand{\Mc}{\mathcal{M}}

\newcommand{\Oc}{\mathcal{O}}

\newcommand{\Rc}{\mathcal{R}}

\newcommand{\Tc}{\mathcal{T}}

\newcommand{\Wc}{\mathcal{W}}

\DeclareMathOperator*{\SumInt}{%
   \mathchoice%
  {\ooalign{$\displaystyle\sum$\cr\hidewidth$\displaystyle\int$\hidewidth\cr}}
  {\ooalign{\raisebox{.14\height}{\scalebox{.7}{$\textstyle\sum$}}\cr\hidewidth$\textstyle\int$\hidewidth\cr}}
  {\ooalign{\raisebox{.2\height}{\scalebox{.6}{$\scriptstyle\sum$}}\cr$\scriptstyle\int$\cr}}
  {\ooalign{\raisebox{.2\height}{\scalebox{.6}{$\scriptstyle\sum$}}\cr$\scriptstyle\int$\cr}}
}

\newcommand{\ie}{i.e.\xspace}

\newcommand{\df}{\textrm{df}}
\newcommand{\nn}{\nonumber}
\newcommand{\diff}{\textrm{d}}
\renewcommand{\th}{{\textrm{th}}}
\newcommand{\pt}{\textrm{pt}}
\newcommand{\pv}{\textrm{pv}}

\newcommand{\beq}{\begin{equation}}
\newcommand{\eeq}{\end{equation}}

\newcommand{\kev}{\ensuremath{{\mathrm{\,ke\kern -0.1em V}}}\xspace}
\newcommand{\mev}{\ensuremath{{\mathrm{\,Me\kern -0.1em V}}}\xspace}
\newcommand{\gev}{\ensuremath{{\mathrm{\,Ge\kern -0.1em V}}}\xspace}
\newcommand{\tev}{\ensuremath{{\mathrm{\,Te\kern -0.1em V}}}\xspace}

\usepackage{mfirstuc} 
\newcommand{\addReviewer}[2]{
  \expandafter\newcommand\csname #1\endcsname[1]{{\sf \color{#2} {#1}:\,##1}}
  \expandafter\newcommand\csname #1cor\endcsname[2]{{\color{#2} {#1}:\,\st{##1}{\sf ##2}}}
  \expandafter\newcommand\csname #1color\endcsname{#2}
}

\usepackage{soul,color}
\definecolor{chromeyellow}{rgb}{1.0, 0.65, 0.0}
\definecolor{DodgeBlue}{rgb}{0.118, 0.565,1.000}
\definecolor{asparagus}{rgb}{0.53, 0.66, 0.42}
\definecolor{cadmiumgreen}{rgb}{0.0, 0.42, 0.24}

\definecolor{jlab_red}{RGB}{192,39,45}
\definecolor{jlab_orange}{RGB}{249,102,0}
\definecolor{jlab_blue}{RGB}{47,122,121}
\definecolor{jlab_green}{RGB}{65,125,10}

\addReviewer{rb}{jlab_blue}
\addReviewer{aj}{jlab_orange}
\addReviewer{mh}{jlab_red}